\documentclass[12pt]{article}
\usepackage{geometry} 
\title{Catastrophic Cracking Courtesy of Quiescent Cavitation\\ Entry\#: 102388}
\author{D. Jesse Daily \and Ken Langley \and Scott L. Thomson \and Tadd T. Truscott\\
\\\vspace{6pt} Department of Mechanical Engineering,\\
Brigham Young University, Provo, Utah, 84602, USA
}
\date{} 

\begin{document}

\maketitle

\begin{abstract}
A popular party trick is to fill a glass bottle with water and hit the top of the bottle with an open hand, causing the bottom of the bottle to break open. We investigate the source of the catastrophic cracking through the use of high-speed video and an accelerometer. Upon closer inspection, it is obvious that the acceleration caused by hitting the top of the bottle is followed by the formation of bubbles near the bottom. The nearly instantaneous acceleration creates an area of low pressure on the bottom of the bottle where cavitation bubbles form. Moments later, the cavitation bubbles collapse at roughly 10 times the speed of formation, causing the bottle to break. The accelerometer data shows that the bottle is broken after the bubbles collapse and that the magnitude of the bubble collapse is greater than the initial impact. This fluid dynamics video highlights that this trick will not work if the bottle is empty nor if it is filled with a carbonated fluid because the vapor bubbles fill with the CO$_2$ dissolved in the liquid, preventing the bubbles from collapsing.  A modified cavitation number, including the acceleration of the fluid (a), vapor pressure (P$_v$), and depth of the fluid column (h), is derived to determine when cavity inception occurs. Through experimentation, visible cavitation bubbles form when the cavitation number is less than 0.5. The experiments, based on the modified cavitation number, reveal that the easiest way to break a glass bottle with your bare hands is to fill it with a non-carbonated, high vapor pressure fluid, and strike it hard.\cite{Daily2013}

\end{abstract}

\bibliographystyle{abbrv}
\bibliography{bib}

\end{document}